\documentclass[aps,pre,twocolumn,showpacs,groupedaddress,
               showkeys,amsmath,amssymb]{revtex4}
\usepackage{graphicx}
\usepackage{bm}
\usepackage{bbm}
\usepackage{epic}
\usepackage{eepic}
\usepackage{pifont}
\usepackage{nicefrac}
\begin{document}
\title{\bf High-Field Low-Frequency Spin Dynamics}
\author{A.M. Farutin, V.I. Marchenko}
\affiliation{P.L. Kapitza Institute for Physical Problems, RAS\\
119334, Kosugina 2, Moscow, Russia}
\email[]{farutin@kapitza.ras.ru, mar@kapitza.ras.ru}
\date{\today}

\begin{abstract}
The theory of exchange symmetry of spin ordered states is extended to
the case of high magnetic field.  Low frequency spin dynamics
equation for quasi-goldstone mode is derived for two cases of
collinear and noncollinear antiferromagnets.
\end{abstract}

\maketitle
Low frequency spin dynamics exists in magnetic materials if
spin-orbital and magnetic dipole-dipole effects are small compared to
the exchange ones. The number of the quasi-goldstone modes and the
spin dynamics equations are determined by the exchange symmetry of a
given spin structure \cite{AM80}. The concept of the exchange
symmetry can be easily extended to the spin structures strongly
deformed by high external magnetic field. In this case there is only
one quasi-goldstone mode. It is related to the invariance of exchange
and Zeeman energies upon the rotation of the spin space around
magnetic field direction on some angle  $\psi$. In this Letter we
derive the low frequency spin dynamics equations for the two
characteristic examples of the antiferromagnets.

From the symmetry point of view there is no difference between the
description of the spin structures with field induced spin
polarization, and those with spontaneous spin polarization
(ferrimagnets). The classification of exchange spin structures is
based on the concept of the exchange space symmetry group $G.$ This
is a group of the charge density symmetry in the exchange
approximation (i.e., without relativistic effects).

The extension of the exchange symmetry concept in our case is
straightforward. Nonetheless one should note that the only effect of
the magnetic field which can be taken into account in the spin
Hamiltonian is the Zeeman term ${-\gamma\hat{{\bf S}}{\bf H}}.$  Here
$\gamma$ is the free electron gyromagnetic ratio, and $\hat{{\bf S}}$
is the total spin operator. The magnetic field effect on the orbital
state of electrons should be neglected. It is easy to check that this
approximation is valid for all known magnets. Namely, the change of
the orbital state remains small even at the spin-flip field.

The dynamics equation for quasi-goldstone degree of freedom $\psi$
can be derived from Lagrange formalism. The kinetic energy is
\begin{equation}
K=\frac{I}{2}{\dot{\psi}}^2. \label{K}\end{equation} At low fields
the moment of inertia $I=\gamma^{-2}\chi,$ where $\chi$ is
magnetic susceptibility (see \cite{AM80}). One can generalize this
result to the high fields. On one hand the magnetization ${\bf M}$
is the derivative of Lagrange function with respect to magnetic
field ${\bf M}=\partial L/\partial{\bf H}$. On the other hand in
the exchange approximation the magnetization is ${\bf
M}=\gamma{\bf S}$. The mechanical spin momentum is oriented along
the magnetic field and its value is $S=\delta L/\delta
\dot{\psi}.$ Let's expand the Lagrange function on small
$\dot{\psi}$
\begin{equation}
L=A+B\dot{\psi}+\frac{I}{2}\dot{\psi}^2+... \label{L}\end{equation}
The two definitions of magnetization give us the following relations
\begin{equation}M=\frac{\partial A}{\partial H}+\frac{\partial B}{\partial
H}\dot{\psi}+...=\gamma(B+I\dot{\psi}+...).\label{M}\end{equation}
The magnetization consists of the static  $M_s=\partial A/\partial
H=\gamma B,$ and dynamic $M_d\approx\partial B/\partial
H\dot{\psi}=\gamma I\dot{\psi}$ parts. From these we see that the
moment of inertia $I=\gamma^{-2}\tilde{\chi},$ where
$\tilde{\chi}$ is the static differential susceptibility.

For the first case consider a collinear antiferromagnets.  At high
magnetic field the antiferromagnetic vector ${\bf L}$ is usually
perpendicular to the field. As it is clear from general results of
the exchange symmetry theory \cite{AM80} the exchange crystal
symmetry group $G$ remains the same as in the absence of magnetic
field. The antiferromagnetic vector transforms according to one of
the one-dimensional representations of the group $G.$ Orientation of
the vector ${\bf L}$ around the field is dictated by the crystal
anisotropy.

Let us introduce Euler angles $\theta, \varphi, \psi$ between unit
orthogonal basis ${\bf a}, {\bf b}, {\bf c}$ in the spin space and
the unit orthogonal basis  ${\bf x}, {\bf y}, {\bf z}$ in the orbital
space. Let us chose the orientations of the basic vectors ${\bf
c}\|{\bf H}$, and ${\bf a}\|{\bf L}.$

The part of the energy which  depends on orientation of the
antiferromagnetic vector ${\bf L}$ is
\begin{equation}
U= \frac{\beta}{2}a^2_z+
\frac{g_{ik}}{2}\partial_i{\psi}\partial_k{\psi}.\label{U1}
\end{equation}
Here we consider the symmetry groups $G$ with one principal axis
$C_3, C_4,$ or $C_6$, $({\bf z}\|C_n).$ The tensor $g_{ik}$ is of an
exchange nature ($g_{zz}=g_\|,$ $g_{xx}=g_{yy}=g_\bot$). The value
$\beta$ corresponds to the relativistic effects of anisotropy. All
the parameters of our theory such as $g_{ik},$ and $\beta$ remain
unknown functions of the magnetic field value. However, the field
orientation dependence of the spin dynamics will be determined.

The energy (\ref{U1}) can be written in the form
\begin{equation}
U= \frac{\beta}{2}\sin^2\theta\cos^2\psi+
\frac{g_{ik}}{2}\partial_i\psi\partial_k\psi. \label{U1a}
\end{equation}
The Lagrange equation corresponding to the kinetic energy (\ref{K})
and the potential energy (\ref{U1})  is
\begin{equation}
I\ddot{\psi}= {\beta}\sin^2\theta\cos\psi\sin\psi+
g_{ik}\partial_i\partial_k\psi. \label{d1}
\end{equation}
The spin wave spectrum $\omega(q)$ is
\begin{equation}
\omega=\sqrt{\omega^2_0+\frac{g_{ik}}{I}q_iq_k}, \label{f1}
\end{equation}
where the gap $\omega_0$ for both possible types of the anisotropy
${\beta>0}$ (easy plane), and ${\beta<0}$ (easy axis) is
\begin{equation}\omega_0=\gamma\sqrt{\frac{|\beta|}{\tilde{\chi}}|}\sin\theta|.
\label{f01}\end{equation} The gap is the longitudinal AFMR frequency.
This statement comes from the expansion (\ref{L}). The second term
$B(H)\dot{\psi}$ is a full derivative for the constant field. But if
the field is time dependent then this term leads to the description
of the longitudinal AFMR.

If the field orientation is close to the main crystal axis
$(\theta=0),$ the gap in the spectrum  goes to zero. The finite gap
value  appears if one takes into account next relativistic terms of
basic plane anisotropy.

Note, that one should add the Dzialoshinskii term
${d(a_xc_y-a_yc_x)=d\sin\theta\sin\psi}$  to the energy (\ref{U1})
for the antiferromagnets with a weak spontaneous ferromagnetism.

For the second case consider the noncollinear antiferromagnet
$Mn_3Al_2Ge_3O_{12}.$ Its AFMR spectrum was studied experimentally
and theoretically at small fields in \cite{PMK86}. The twelve
sublattices spin structure of $Mn_3Al_2Ge_3O_{12}$ is described in
terms of two antiferromagnetic vectors ${\bf L}_1\bot{\bf L}_2$ which
transform according to two-dimensional representation $E_u$ of the
crystal class $O_h.$

As observed in \cite{PMK86} the spin plane tends to become
perpendicular to the field. So, the most probable state of the spin
structure at high fields is ${\bf L}_1,{\bf L}_2\bot{\bf H}.$ Then,
according to \cite{AM80} the exchange space group remains $O_h^{10}$.

At high fields the anisotropy  effects are described to a first
relativistic approximation by the same invariant
\begin{equation}
U_a=\beta\left\{\sqrt{3}(a_z^2-b_z^2)+2(a_xb_x-a_yb_y)\right\}
\label{U2} \end{equation} as in the small field limit (see
\cite{PMK86}). Let us align the basic vectors  ${\bf x}, {\bf y},
{\bf z}$ in the orbital space along fourth-order symmetry axes. The
basis in the spin space is chosen as ${\bf c}\|{\bf H}$, ${\bf
a}\|{\bf L}_1$, and ${\bf b}\|{\bf L}_2.$ In terms of Euler angles
the anisotropy energy (\ref{U2}) can be rewritten as $\beta
f\cos2(\psi-\psi_0),$ where parameters $f$, and $\psi_0$ are the
functions of the field orientation
\begin{gather*}
f=\sqrt{ \cos^2{2\varphi}(1+\cos^2\theta)^2+(\sqrt{3}\sin^2\theta +
2\sin{2\varphi}\cos\theta)^2},\\
\tan2\psi_0=\frac{\cos{2\varphi}(1+\cos^2\theta)}{\sqrt{3}\sin^2\theta
+ 2\sin{2\varphi}\cos\theta}.
\end{gather*}
The magnon spectrum here is given by the general formula (\ref{f1}),
where $g_{ik}=g\delta_{ik},$ and
\begin{equation}\omega_0=2\gamma\sqrt{\frac{2|\beta|f}{\tilde{\chi}}}.
\label{f02}\end{equation}

The anisotropy term  (\ref{U2}) is the main relativistic correction
to the exchange energy of $Mn_3Al_2Ge_3O_{12}.$ This fact leads to a
new relation between static and dynamic characteristics. Consider
anisotropic correction to the magnetization. The ground state
anisotropy energy is $U_a=-4|\beta|$ if the magnetic field is
oriented along [001] axis ($\theta=0$), and $U_a=-16|\beta|/3$ if it
is oriented along [111] ($\cos\theta=1/\sqrt{3}, \varphi=\pi/4$). It
gives an opportunity to find the function $\partial\beta/\partial H.$
Namely, \begin{equation}\frac{\partial|\beta|}{\partial
H}=\frac{3}{4}\left(M_{[111]}-M_{[001]}\right).\label{MM}\end{equation}

The obtained results can be applied to any usual magnetics, as
well as to tensors spin structures \cite{FM05}. In  all cases
there are additional relations of type (\ref{MM}). But, the
situations can be more complicated. It may become necessary to
take into account small deviations of the magnetization from the
field orientation.

In conclusion, we have presented the exchange symmetry theory of
spin ordered states strongly deformed by high magnetic field. Spin
dynamics equation is derived for quasi-goldstone mode which is the
rotation of spin space around the field direction. The dispersion
of the spin waves is determined. A new relation between static and
dynamic characteristics is found.

We thank A.F. Andreev, L.A. Prozorova, and A.I. Smirnov for helpful
discussions. The work is supported by RFBR Grants No. 04-02-17294,
06-02-16509, and 06-02-17281, RF President Program, and Landau
Scholarship (A.F.) from Forschungszentrum Juelich.


\begin{thebibliography}{9}
\bibitem{AM80} A.F. Andreev and V.I. Marchenko,
Sov. Phys. Uspekhi. {\bf 23}, 21 (1980)
\bibitem{PMK86} L.A. Prozorova, V.I. Marchenko, and Yu.V. Krasnyak,
JETP Lett. {\bf 41}, 637 (1986)
\bibitem{FM05} A.M. Farutin and V.I. Marchenko,
JETP. {\bf 100}, 977 (2005)
\end{thebibliography}
\end{document}